\documentclass[12pt]{article}

\usepackage[linktocpage=true,plainpages=false]{hyperref}
\usepackage{color}
\usepackage{amsmath,amssymb}
\usepackage{cite}
\usepackage{graphicx}
\usepackage{array}
\usepackage{multicol}
\usepackage[T1]{fontenc}
\usepackage[labelsep=period]{caption}

\definecolor{linkcolor}{rgb}{0.6,0,0}
\definecolor{citecolor}{rgb}{0,0.6,0}
\definecolor{urlcolor}{rgb}{0,0,0.9}
\hypersetup{colorlinks, linkcolor={linkcolor},citecolor={citecolor}, urlcolor={urlcolor}}

\newcommand{\dd}{\partial}
\newcommand{\de}{\delta}
\newcommand{\m}{\mu}
\newcommand{\n}{\nu}
\newcommand{\ls}{\left(}
\newcommand{\rs}{\right)}
\newcommand{\al}{\alpha}
\newcommand{\na}{\nabla\!}
\newcommand{\str}[1]{\mathrel{\mathop{\longrightarrow}\limits_{#1}}}
\newcommand{\ff}{\varphi}
\newcommand{\te}{\theta}
\newcommand{\be}{\beta}
\newcommand{\ga}{\gamma}
\newcommand{\om}{\omega}
\newcommand{\p}{\bot}
\newcommand{\po}{{\Pi_{\!\!\bot}}}
\newcommand{\ta}{\tau}
\newcommand{\sh}{\sinh}
\newcommand{\ch}{\cosh}

\newcommand{\disn}[2]{$$\displaylines{\refstepcounter{equation}%
            \label{#1}\hskip 1em minus 1em #2\hfilneg}$$}
\newcommand{\nom}{\hfil\hskip 1em minus 1em (\theequation)}
\newcommand{\no}{\hfil \hskip 1em minus 1em\phantom{(\theequation)}%
            \hfilneg\cr\hfilneg\hskip 1em minus 1em\hfil}
\newcommand{\ns}{\hfill\cr\hfill}

\begin{document}
\title{Relation between quantum effects\\ in General Relativity and embedding theory}
\author{S.~A.~Paston\thanks{E-mail: paston@pobox.spbu.ru}
\\
{\it Saint Petersburg State University, Saint Petersburg, Russia}
}
\date{\vskip 15mm}
\maketitle

\begin{abstract}
We present results relevant to the relation between quantum effects in a Riemannian space and on the
surface appearing as a result of its isometric embedding in a flat space of a higher dimension. We discuss
the mapping between the Hawking effect fixed by an observer in the Riemannian space with a
horizon and the Unruh effect related to an accelerated motion of this observer in the ambient space. We
present examples for which this mapping holds and examples for which there is no mapping.
We describe the general form of the hyperbolic embedding of the metric with a horizon smoothly covering
the horizon and prove that there is a Hawking into Unruh mapping for this
embedding. We also discuss the possibility of relating two-point functions in a Riemannian space and
the ambient space in which it is embedded. We obtain restrictions on the geometric parameters of the
embedding for which such a relation is known.
\end{abstract}

\newpage

\section{Introduction}
In accordance with Einstein's General Relativity, gravitational interaction is described in the framework
of the assumption that our four-dimensional space-time is a Riemannian space, i.~e., a manifold on which
the metric $g_{\m\n}(x)$ and the connection corresponding to it are defined (here and hereafter, $\m,\n,\ldots=0,1,2,3$,
and we use the signature $(+---)$). Although there is currently no satisfactory quantum gravity theory, we
can study the influence of the gravity on quantum effects in the case of not too high energies, considering
quantized fields in a Riemannian space (see, e.g., \cite{birrell,grib}). The most well-known nontrivial examples of such
an influence can include the Hawking effect \cite{hawking75} (radiation from black holes) and the Unruh effect
\cite{unruh} (the
observation of thermal radiation by a uniformly accelerated observer).

It follows from the Janet-Cartan theorem generalized to the pseudo-Riemannian case by Friedman \cite{fridman61}
that an arbitrary $n$-dimensional pseudo-Riemannian space can be at least locally embedded isometrically
in a flat pseudo-Euclidean space $R^{m,N-m}$ with the number of dimensions $N\geqslant n(n+1)/2$. In this case, the
pseudo-Riemannian space is set into correspondence with the surface described by the embedding function
$y^a(x^\m)$, with the metric on the surface,
\disn{7.1}{
g_{\m\n}=(\dd_\m y^a) (\dd_\n y^b) \eta_{ab}
\nom}
induced by the flat metric $\eta_{ab}$ of the ambient space (here and hereafter, the superscripts $a,b,\ldots$ take $N$
values). This allows passing to the description of gravity in the form of a theory corresponding to our
space-time of a four-dimensional surface in the flat ten-dimensional ambient space. In such an approach,
which is usually called the embedding theory, instead of the metric $g_{\m\n}(x^\al)$ as the independent variable
defining gravity, the embedding function $y^a(x^\al)$ is chosen.

Such an approach to the description of gravity, which is similar to that used in string theory, was first
proposed by Regge and Teitelboim \cite{regge}. Different variants of such an approach were later studied in rather
many papers, for example, in
\cite{deser,pavsic85let,tapia,maia89,bandos,estabrook1999,davkar,bustamante,statja18,faddeev,statja26,willison}.
The idea of using the embedding appeared independently in \cite{pavsic85let,estabrook1999}.
We note that unlike the recently popular brane theory \cite{lrr-2010-5}, the ambient space is flat in embedding theory,
and there is no gravity in it. Therefore, unlike brane theory, embedding theory is an alternative description
of gravity in terms of the embedding function $y^a(x^\al)$ and has the form of a certain theory in a flat space-time
in this case, which can be useful for efforts to quantize gravity. There also exists a possibility of formulating
embedding theory in the form of a certain field theory in a flat ambient space: splitting theory \cite{statja25}, which
brings together the description of gravity in such an approach and the description of other fundamental
interactions for which the correct quantum formulation is constructed.

Regarding our space-time as a four-dimensional surface in a flat space, we can formulate a problem of
the relation between the quantum effects in space-time as the Riemannian space (mentioned above) and the
effects appearing on the surface because of the presence of quantum fields in the flat ambient space. It turns
out that such a relation exists, at least for certain types of embeddings of a certain class of Riemannian
spaces, including some that are physically interesting solutions of the Einstein equations.

As mentioned by Deser and Levin \cite{deserlev98,deserlev99} (we also note \cite{culetu}, which was published at the same time
but is less known), the black hole radiation predicted by Hawking can be identified with the Unruh radiation
observed in the ambient space. For this, we must consider a certain embedding of the black hole metric (i.~e.,
the four-dimensional surface with such a metric in a flat ambient space) for which an observer that was at
rest at a fixed distance from the black hole turns out to be moving uniformly accelerated in the ambient
space. Such a correspondence between the Hawking and Unruh effects ("Hawking into Unruh mapping")
was discovered for many physically
interesting metrics. In detail in Secs.~3-7, we discuss the cases in which this mapping occurs and
those in which there is no such mapping. Following \cite{statja36}, we also formulate and prove the statement
describing the conditions on the metric and the embedding function under which the mapping holds.

Another example of the relation between quantum effects in a Riemannian space and on a surface
corresponding to it and embedded in an ambient flat space is the possibility of relating the two-point
Wightman functions (or the Green's functions) in the Riemannian and ambient spaces. This can be done
for surfaces corresponding to a so-called warped geometry. The corresponding analyses were performed for
the Wightman functions in \cite{hep-th0003098} and for the Green's function in \cite{1305.6933,1310.2939}.
In Sec.~8, we discuss the class of surfaces for which this approach is applicable.

\section{Hawking and Unruh effects}
In the most general sense, the Hawking effect \cite{hawking75} is that in a space-time with an event horizon, an
observer outside the horizon observes a flux of particles with a thermal distribution. The nontriviality
of the causal structure of the space-time with a horizon in which the quantum field theory is considered
underlies this quantum effect. Because of this nontriviality, a state vector that is the vacuum state from
the standpoint of field operators on the initial surface in the past, being expanded in terms of the basis
of states corresponding to field operators at the final surface in the future (including the infinitely distant
surface and the event horizon), corresponds to the presence of radiation with a thermal spectrum.
The radiation temperature determined by the observer (the Hawking temperature) is expressed by the
formula
\disn{n1}{
T_H=\frac{k}{2\pi}
\nom}
in terms of the so-called "surface gravity" $k$, which is defined for a given observer as follows
(see, e.g., \cite{townsend}).
Let $\xi$ be the Killing timelike vector normalized such that its value $\xi_0$ at the observer location satisfies the
condition $\xi_0^2=1$. Let the space-time contain a lightlike hypersurface $\cal N$ on which $\xi$ is orthogonal to all
vectors tangent to $\cal N$ (in this case, the vector $\xi$ is the lightlike vector tangent to $\cal N$).
Then $\cal N$ is called the
Killing horizon. At points of the surface $\cal N$, we have the relation
\disn{2}{
\xi^\al\na_\al\xi^\m=k\xi^\m,
\nom}
defining the surface gravity $k$ (here, $\na_\al$ denotes the covariant derivative). From the physical standpoint, $k$
corresponds to the four-dimensional acceleration of a particle that is at rest near the horizon in the static
reference system (the Killing vector $\xi$ is tangent to its world line), which is rescaled with the redshift taken
into account for an observer located at some point outside the horizon.

For three-dimensional infinite spaces for which $\xi^2$ is bounded at three-dimensional infinity, for example,
for the majority of black holes, it is usually assumed that the observer is located at infinity. The Killing
vector is then normalized by the condition $\xi^2\str{r\to\infty}1$.
But when considering spaces with finite threedimensional
sizes, for example, de Sitter (dS) spaces, it is necessary to analyze the Hawking effect from the
standpoint of an observer located at a finite distance.

The Unruh effect \cite{unruh} is that in the Minkowski space, in a state that is the vacuum state from the
standpoint of an inertial observer, an observer moving with a constant acceleration $w$ sees thermal radiation
with the Unruh temperature
\disn{2.1}{
T_U=\frac{w}{2\pi}.
\nom}
To see this, a certain field in the Rindler coordinates \cite{rindler} $t,\rho,x^{2,3}$
comoving for the uniformly accelerated
observer and related to the Lorentz coordinates $x^\m$ by the expressions
\disn{3}{
x^0=\rho\sh t,\qquad x^1=\rho\ch t
\nom}
can be quantized. If the basis of states corresponding to such a quantization is constructed, then the
quantum theory vacuum given in the Lorentz coordinates can be expanded in terms of it, and the thermal
distribution of so-called "Rindler" particles with temperature (\ref{2.1}) can be observed (the Bogoliubov transformation
is standardly used for this; see, e.g., \cite{birrell}). Such an approach is very similar to that used to describe
Hawking radiation, which leads to the well-known trivial Hawking into Unruh mapping described in the next section. The presence of a horizon for the Rindler observer plays an important
role in this case.

But because the procedure for observing the Unruh radiation was not described in the framework of
such an approach, it does not offer the possibility of unambiguously answering the question whether this
radiation is real. The interpretation problem turns out to be especially complicated when considering not
rectilinear accelerated motion but uniform motion along a circle during which the absolute value of the
acceleration remains constant. There is no horizon for such a motion, and the vacuum turns out to coincide
with the that of the initial reference system \cite{letaw80}, i.~e., it should be assumed that there is no Unruh effect.
Nevertheless, as noted in \cite{akhmedov07}, it is strange to assume that there is no Unruh effect if there is no horizon
because we can consider an observer moving rectilinearly with the same acceleration only during a finite
time interval before and after which it is inertial. There is no horizon for such an observer, but to assume
that at the instant of his linearly accelerated motion, the observer does not see Unruh radiation means
assuming that the presence or the absence of the radiation depends nonlocally on the past or future of
the observer. It seems more reasonable to assume that the parameters of the observed radiation must be
determined by local characteristics of the observer trajectory.

The description of the Unruh effect by considering the Unruh-DeWitt detector moving along a given
trajectory and interacting with the quantum field in the space-time is more consistent from the physical
standpoint. The Unruh-DeWitt detector (see, e.g., \cite{birrell} for more details) is a point object described by the
monopole moment operator $m(\ta)$ of the detector and can transform from its ground to the excited state
with the energy $E$. As the Hamiltonian of interaction between the detector and the quantum field (for
example, the scalar field $\ff(x^\m)$), we take the simplest expression
\disn{n2.1}{
H_I=\int d\ta\,g\, m(\ta)\ff(x^\m(\ta))
\nom}
where $\ta$ is the proper time of the detector. The probability of the transition to the state with the energy
$E$ can then be calculated in the lowest order of the perturbation theory in the coupling constant $g$, and the
resulting spectrum can be found.

Most often, stationary trajectories of the detector motion, to which both rectilinear uniformly accelerated
motion and the motion along a circle are related, are considered. In \cite{letaw81}, stationary motions were
completely classified, and the spectra corresponding to them were calculated. It is interesting that in that
paper and some others, the term "Unruh radiation" was not used for trajectories without horizons, and only
the vacuum excitation spectrum of the detector was mentioned. Unlike the case of rectilinear uniformly
accelerated motion, for all other stationary trajectories, the spectrum turns out to be nonthermal. It is
interesting that in \cite{akhmedov07}, the Unruh effect for motion along a circle was identified with the experimentally
observed Sokolov-Ternov effect. A discussion of various aspects of the Unruh effect can also be found in
the literature recently cited in \cite{abdolrahimi}.

The interaction Hamiltonian in form (\ref{n2.1}) allows finding only the excitation spectrum appearing as a
result of the detector-quantum-field interaction at all instants of the detector proper time. To obtain
the "instantaneous" spectrum that, as can be expected, must correspond to the observed radiation, the
interaction must be modified somehow. This was done in \cite{barbado2012} by introducing a dependence of the coupling
constant $g$ on the time $\ta$ in formula (\ref{n2.1}) by turning the interaction on and off at certain instants. It was
shown that in the case of arbitrary one-dimensional motion with a slowly varying acceleration (i.~e., satisfying
the condition $|\dot w| \ll w^2$), there exists a thermal spectrum with standard Unruh temperature (\ref{2.1}).

\section{Trivial Hawking into Unruh mapping}\label{triv}
We first describe the well-known trivial Hawking into Unruh mapping, which
occurs when considering the motion of a uniformly accelerated observer in the Minkowski space. Such
an observer is at rest in the Rindler coordinates $t,\rho,x^{2,3}$ defined by formulas (\ref{3}) or, if the lightlike
coordinates $x^\pm=x^0\pm x^1$ are used, by the formulas
\disn{3.4}{
x^+=\rho e^t,\qquad x^-=-\rho e^{-t}.
\nom}
Translation of the Rindler coordinates with respect to the time $t$ corresponds to a Lorentz boost in the
Minkowski space. The corresponding Killing vector $\xi^\m$ normalized to unity at $\rho=\rho_0$ has the components
\disn{3.5}{
\xi^+=\frac{1}{\rho_0}\rho e^t=\frac{1}{\rho_0}x^+,\qquad \xi^-=\frac{1}{\rho_0}\rho e^{-t}=-\frac{1}{\rho_0}x^-,
\qquad \xi^{2,3}=0.
\nom}
The Killing horizon $\cal N$, which is the surface $(x^+=0\bigcup x^-=0)\bigcap x^1\ge 0$, exists for it.
Therefore, the surface
gravity can be determined, and we can assume that the Hawking effect occurs for the Rindler observer.

Taking into account that in the Lorentz coordinates, the covariant derivative $\na_\al$ in the Minkowski
space reduces to the ordinary derivative $\dd_\al$, it is easy to calculate the left-hand side of Eq.~(\ref{2})
at a certain horizon point, for example, at a point where $x^-=0,x^+>0$:
\disn{3.6}{
\xi^\al\na_\al\xi^\m=\xi^+\dd_+\xi^\m=\de^\m_+\frac{x^+}{\rho_0^2}=\frac{1}{\rho_0}\xi^\m.
\nom}
We hence find the surface gravity from the standpoint of the observer with the coordinate $\rho_0$:
\disn{3.7}{
k=\frac{1}{\rho_0}.
\nom}
It is obvious that this value coincides with the constant acceleration $w$ of this observer, and this means that
we can say that Hawking temperature (\ref{1}) and Unruh temperature (\ref{2.1}) coincide for the Rindler observer in
the Minkowski space, i.~e., this is the simplest case of the Hawking into Unruh mapping.

\section{Hawking into Unruh mapping at using embeddings}
As it turned out, the Hawking into Unruh mapping analogous to the mapping
described in the preceding section occurs if a certain type of isometric embedding in a flat
ambient space for a rather wide class of metrics is considered. It was first revealed
in \cite{deserlev97,deserlev98,deserlev99} for de Sitter (dS)
and anti-de Sitter (AdS) metrics, for the Schwarzschild metric with its dS and AdS generalizations, for the
Reissner-Nordstrom charged black hole metric, and also for the Banados-Teitelboim-Zanelli metric corresponding
to the three-dimensional black hole. It was noted that for the considered embeddings (smoothly
covering the horizon) of these metrics, the temperature $T_H$ corresponding to the Hawking effect due to
the presence of the horizon coincides with the temperature $T_U$ of the Unruh effect due to the uniformly
accelerated motion of the observer from the standpoint of the ambient space.
For example, for the Schwarzschild metric
\disn{0.1}{
ds^2=\ls 1-\frac{2m}{r}\rs dt^2-\frac{dr^2}{1-\frac{2m}{r}}-r^2\ls d\te^2+\sin^2\te\, d\ff^2\rs
\nom}
the Fronsdal embedding \cite{frons} in a six-dimensional flat space with the signature $(+-----)$:
\disn{1}{
y^0=\tau\sqrt{1-\frac{2m}{r}}\,\sh\ls\frac{t}{\tau}\rs,\quad
y^1=\pm\tau\sqrt{1-\frac{2m}{r}}\,\ch\ls\frac{t}{\tau}\rs,\quad \text{at}\quad r>2m,\no
y^0=\pm\tau\sqrt{\frac{2m}{r}-1}\,\ch\ls\frac{t}{\tau}\rs,\quad
y^1=\tau\sqrt{\frac{2m}{r}-1}\,\sh\ls\frac{t}{\tau}\rs,\quad \text{at}\quad r<2m,\no
y^2=\int dr\sqrt{\frac{2m}{r-2m}\ls 1-\frac{m\tau^2}{2r^3}\rs},\no
y^3=r\,\cos\te,\quad
y^4=r\,\sin\te\,\cos\ff,\quad
y^5=r\,\sin\te\,\sin\ff
\nom}
was used. This embedding smoothly covers the horizon at $\tau=4m$. It is easy to see that for it, the time
lines are hyperbolas and correspond to uniformly accelerated motion in the ambient space for $r>2m$. It
was found that the Hawking and Unruh temperatures coincide precisely for a single value of the embedding
parameter $\tau=4m$, for which the component $y^2$ in (\ref{1}) does not contain singularities and the embedding
smoothly covers the event horizon.

A similar Hawking into Unruh mapping for many other metrics was subsequently
found in many papers. The corresponding results were obtained for different types of black holes
(including those with other than four dimensions), black strings, and wormholes (see
\cite{kim00,lemos,arXiv:1012.5709,hep-th/0103036,gr-qc/0303059,arXiv:1311.0592}
and the references therein). The cases where the detector moves along a circle \cite{gr-qc/0409107}
or falls freely \cite{arXiv:0805.1876} were also
discussed. In these cases, the radiation spectrum was not exactly thermal. It was shown in \cite{arXiv:0901.0466}
that the coincidence is also preserved under corrections to the Hawking and Unruh temperatures. It was noted
in \cite{Banerjee} that the mapping occurs if the embedding of only the $(t-r)$ sector of the Riemannian space is
considered. It is important that in all the cases, hyperbolic embeddings covering the horizon were used,
their time dependences were similar to (\ref{1}) (they are also called the Fronsdal-type embeddings), and the
time lines in them are hyperbolas in the ambient space.

The approach in which the thermodynamic properties of spaces with horizons are analyzed using
isometric embeddings in a flat ambient space is usually called GEMS
(Global Embedding Minkowskian Space-time). We note that this term is slightly
inexact: the embeddings used in its framework are not in fact always global, i.~e., smooth at all nonsingular
points of the Riemannian space. For example, when analyzing the Hawking into Unruh mapping for the metric of the Reissner-Nordstrom charged black hole in \cite{deserlev99}, an embedding smoothly
covering the external horizon was used, but it cannot be smoothly continued past the internal horizon
(unlike embeddings proposed in \cite{statja30}). Hence, under "globality" in the term GEMS, only the fact that the
embedding smoothly covers the observed region of the space with an horizon is meant, and whether the
smoothness is preserved past the horizon is assumed to be unimportant.

\section{Examples of the absence of the mapping}
Despite many examples of embeddings in which we have the Hawking into Unruh mapping described in the preceding section,
counterexamples also exist. We describe them following \cite{statja34}.

In particular, relatively recently constructed embeddings of the Schwarzschild metric \cite{davidson,statja27}
(including
its dS generalization \cite{statja32}) and the Reissner-Nordstrom metric \cite{statja30} are such counterexamples. These
new embeddings are of three types, and all of them are not hyperbolic, i.~e., their time lines are not hyperbolas.
To clarify their structure, we present the explicit forms of embeddings of all three types for the
Schwarzschild metric (the embedding structures of the other mentioned metrics are analogous): the cubic
embedding
 \disn{4}{
\begin{array}{l}
\displaystyle y^0=\frac{\xi^2}{6}t'^3+\ls 1-\frac{m}{r}\rs t'+u(r),\\[1em]
\displaystyle y^1=\frac{\xi^2}{6}t'^3-\frac{m}{r}t'+u(r),\\[1em]
\displaystyle y^2=\frac{\xi}{2}t'^2+\frac{1}{2\xi}\ls 1-\frac{2m}{r}\rs,
\end{array}
\nom}
the exponential embedding
\disn{5}{
\begin{array}{l}
\displaystyle y^0=\frac{m}{\al\sqrt{r_c r}}\ls e^{\al t'+u(r)}-\frac{r-r_c}{2m}\,e^{-\al t'-u(r)}\rs,\\[1em]
\displaystyle y^1=\frac{m}{\al\sqrt{r_c r}}\ls e^{\al t'+u(r)}+\frac{r-r_c}{2m}\,e^{-\al t'-u(r)}\rs,\\[1em]
\displaystyle y^2=\hat\ga t'
\end{array}
\nom}
and the spiral embedding
 \disn{6}{
\begin{array}{l}
\displaystyle y^0=t',\\[0em]
\displaystyle y^1=\frac{(6m)^{3/2}}{\sqrt{r}\quad}\,\sin\ls\frac{t'}{3^{3/2}2m}-\sqrt{\frac{2m}{r}}\,\bigg( 1+\frac{r}{6m}\bigg)^{3/2}\,\rs,\\[1.2em]
\displaystyle y^2=\frac{(6m)^{3/2}}{\sqrt{r}\quad}\,\cos\ls\frac{t'}{3^{3/2}2m}-\sqrt{\frac{2m}{r}}\,\bigg( 1+\frac{r}{6m}\bigg)^{3/2}\,\rs.
\end{array}
\nom}
Here, $t'$ is the time of some falling coordinates related to the Schwarzschild time by $t'=t+h(r)$, the
constants $\xi,\al,\hat\ga,r_c$ are the embedding parameters, and the components $y^3,y^4,y^5$ and the
signature of the ambient space coincide with those used for embedding (\ref{1}).

The time lines for all the three embeddings are not hyperbolas. They are stationary trajectories, and
an analogue of the Unruh effect for them was studied in \cite{letaw81}. For example, the spectrum for a detector
moving in the ambient space along the time lines of embedding (\ref{4}) was found exactly. It has the form
 \disn{7}{
S(E)=\frac{E^2}{8\pi^2\sqrt{3}\,\xi^2}\ls 1-\frac{2m}{r}\rs^2\exp\ls -\frac{\sqrt{12}}{\xi^2}\ls 1-\frac{2m}{r}\rs E\rs,
\nom}
and is not thermal. The spectrum for the time lines of embedding (\ref{5}) was studied numerically in \cite{letaw81} and
was also recently analyzed in \cite{abdolrahimi}. As in the preceding case, the spectrum is not thermal.
The time lines of
embedding (\ref{6}) correspond to detector motion along a circle. The analogue of the Unruh effect in this case
has been frequently discussed in the literature. The spectrum is also not thermal in this case. Moreover,
as can be easily seen from (\ref{6}), the circle radius tends to zero as $r\to\infty$ (for a fixed angular velocity),
and the detector turns out to be at rest in the chosen coordinate system of the ambient space. This means
that there must indeed be no Unruh effect in this limit. Naturally, such a case is related to the fact that
embedding (\ref{6}) is asymptotically flat, i.~e., the surface corresponding to it tends to the plane as $r\to\infty$.
Precisely such an embedding turns out to be most natural from the standpoint of embedding theory as a
description of gravity (see the discussion in \cite{statja27}).

Hence, for all three new six-dimensional embeddings of the Schwarzschild metric, there is no Hawking into Unruh mapping. An analogous case also occurs for three new embeddings of
the Reissner-Nordstrom metric found in \cite{statja30} and for two new embeddings of the Schwarzschild-de Sitter
metrics found in \cite{statja32} because their structures are completely analogous to embeddings (\ref{4})-(\ref{6}).

Another class of examples of the absence of the Hawking into Unruh mapping can be obtained easily by the trivial isometric bending (such as the plane bending to form a part
of a cylinder) of a flat ambient space containing a surface for which the mapping existed before the
bending. For example, if a standard hyperboloid is bent, then we can obtain a nonstandard embedding for
the dS metric in the flat space with the signature $(+-----)$ of the form
 \disn{5.8}{
\begin{array}{l}
y^0=R\, \sinh(\sinh t),\\
y^1=R\, \cosh(\sinh t),\\
y^2=R\, \cosh t\,\cos\chi,\\
y^3=R\, \cosh t\,\sin\chi\,\cos\te,\\
y^4=R\, \cosh t\,\sin\chi\,\sin\te\,\cos\ff,\\
y^5=R\, \cosh t\,\sin\chi\,\sin\te\,\sin\ff,
\end{array}
\nom}
The time lines for it are not hyperbolas (and they are even nonstationary trajectories). This means that
the spectrum corresponding to such a motion is not thermal, and there is no mapping.

Another example is the result of bending the four-dimensional plane described by the embedding
function
 \disn{5.9}{
\begin{array}{l}
y^0=\al^{-1} \sinh \al t,\\
y^1=\al^{-1} \cosh \al t,\\
y^2=r\cos\te,\\
y^3=r\sin\te\cos\ff,\\
y^4=r\sin\te\sin\ff
\end{array}
\nom}
with the signature of the ambient space $(+----)$. It is easy to verify that the metric coinciding with
that of the Minkowski space corresponds to it, i.~e., function (\ref{5.9}) specifies a nontrivial embedding of the
Minkowski space in a flat five-dimensional ambient space. If an Unruh-DeWitt detector moves along the
time line of such an embedding, then it observes no radiation from the standpoint of the intrinsic geometry
because it moves along the inertial trajectory in the four-dimensional Minkowski space. But the time line
is a hyperbola in the ambient space, the motion of the detector is linearly accelerated, and the standard
Unruh effect must occur. Hence, there is no mapping in this example.

If the examples for which the presence or the absence of the Hawking into Unruh mapping is verified are analyzed, then it can be seen that it exists in the cases where the metric has a
horizon and the embedding is hyperbolic and smoothly covers the horizon. In Secs.~6 and 7, following \cite{statja36},
we show that the mapping indeed always occurs under these conditions.

\section{General form of the hyperbolic embedding\\ with a horizon}\label{hyp}
We first formulate the general form that has a hyperbolic embedding of a metric with a horizon.
Because we are interested in metrics for which Hawking radiation exists, we assume that there is a timelike
Killing vector $\xi$ and the Killing horizon exists (see, e.g., \cite{townsend}), i.~e.,
a lightlike hypersurface $\cal N$ such that
the vector $\xi$ is tangent to it and $\xi^2=0$ on it. Otherwise, the metric is assumed to be arbitrary, i.~e., it
has no additional symmetry and, general speaking, is not a solution of the Einstein equations with any
definite matter. We choose the coordinates $t,\rho,x^2,x^3$ such that the vector $\xi$ is tangent to the lines
of the time $t$ and the coordinate $\rho$ becomes zero on the hypersurface $\cal N$, i.~e.,
$\xi^2\str{\rho\to0}0$. We assume that
an observer (for example, the Unruh-DeWitt detector) is at rest in the chosen coordinate system and has
constant coordinates $\rho=\rho_0>0,x^2,x^3$ and that $\xi^2>0$ for $0<\rho\le \rho_0$.

The existence of a timelike Killing vector means that the metric has a symmetry with respect to the
one-dimensional group of shifts in time. To construct the surface $\cal M$ that is the embedding of the considered
metric and also has the symmetry with respect to this group, we must consider the representations of the
group of translations whose matrices correspond to the transformations of the Poincare group of the ambient
space (see the details of the method for constructing symmetric embeddings in \cite{statja27}). In one of the simplest
cases, these matrices correspond to the Lorentz boosts in the ambient space, i.~e., a shift in $t$ is equivalent
to a boost. In this case, time lines in the ambient space turn out to be hyperbolas, and embeddings of
such type are said to be hyperbolic (or, as noted above, embeddings of the Fronsdal type). An arbitrary
embedding of the hyperbolic type can be written in the form
\disn{8}{
y^0=\rho\sh(\al t),\qquad
y^1=\rho\ch(\al t),\qquad
y^I=y^I(\rho,x^2,x^3).
\nom}
in the range $0\le \rho\le \rho_0$ (where $\xi^2>0$) if we use the arbitrariness in choosing
the coordinates $\rho,x^2,x^3$. Here and hereafter, $I=2,\ldots,N-1$, and $\al$ is an arbitrary positive
constant introduced for generality.
We note that the components $y^I$ of the embedding function $y^a(x^\m)$ depend on the coordinates $\rho,x^2,x^3$
and not on $t$. When constructing embeddings of concrete metrics below, we must substitute general
embedding form (\ref{8}) in Eq.~(\ref{7.1}) and seek its solution. It is usually impossible to do this without any
additional symmetry, but we can perform the general analysis that is interesting to us.

We normalize the Killing vector $\xi$ corresponding to shifts in the time $t$ such that it is unity at the
observation point $\rho=\rho_0$. In the coordinates $t,\rho,x^2,x^3$ its components then have the form
\disn{9}{
\xi^\m=\de^\m_0\frac{1}{\sqrt{g_{00}(\rho_0)}}=\de^\m_0\frac{1}{\al \rho_0},
\nom}
and in the ambient space, the vector $\xi^a=\xi^\m\dd_\m y^a$ with the components
\disn{10}{
\xi^0=\frac{\rho}{\rho_0}\ch(\al t),\qquad
\xi^1=\frac{\rho}{\rho_0}\sh(\al t),\qquad
\xi^I=0
\nom}
corresponds to it.

We assume that the embedding smoothly covers the horizon, i.~e., the considered four-dimensional
surface is smooth at $\rho=0$. Because the coordinate $t$ turns out to be singular at the horizon (for example,
like the time of the Schwarzschild coordinates for a black hole), the embedding function is not necessarily
smooth if this coordinate is used, and the smoothness therefore cannot be determined directly. For this,
we note that the vector $\xi^a$, remaining tangent to the surface $\cal M$, has strongly different directions in an
arbitrarily small neighborhood of the point $y^0=y^1=0$, but they are in the $y^0,y^1$ plane. This is consistent
with the smoothness of the surface $\cal M$ only in the case where the entire $y^0,y^1$ plane is tangent to $\cal M$ at
the point $y^0=y^1=0$. Consequently, the quantities $y^0,y^1$ can be used as coordinates on $\cal M$ in a
neighborhood of $y^0=y^1=0$, and the smoothness of $\cal M$ guarantees the smoothness of the embedding
function written in terms of these coordinates. If the quantities $y^0,y^1,x^2,x^3$ are used as coordinates
on the surface $\cal M$ in this neighborhood, then embedding function (\ref{8}) (in the range where this formula is
applicable, i.~e., for ${y^1}^2-{y^0}^2\ge0$) becomes
\disn{11}{
y^0=y^0,\qquad
y^1=y^1,\qquad
y^I=y^I\ls\sqrt{{y^1}^2-{y^0}^2},x^2,x^3\rs.
\nom}
Its smoothness with respect to $y^0,y^1$ means that the function $y^I(\rho,x^2,x^3)$ depends smoothly on ${\rho}^2$. It
hence follows that in a neighborhood of the horizon, to which the set of points satisfying the equation
\disn{11.1}{
y{^1}^2-{y^0}^2=0
\nom}
corresponds in the ambient space, the expansion
\disn{12}{
y^I=f^I(x^2,x^3)+\ls{y^1}^2-{y^0}^2\rs h^I(x^2,x^3)+O\ls\ls{y^1}^2-{y^0}^2\rs^2\rs
\nom}
holds. The validity of this expansion is therefore a criterion for the considered embedding to cover the
horizon smoothly.

\section{Proof of the existence of the mapping}\label{uslov}
We prove the following statement.

\textit{\textbf{Statement.} The Hawking into Unruh mapping occurs for a hyperbolic
embedding (that smoothly covers the horizon and can always be written in form (\ref{8})) of an arbitrary metric
with a Killing timelike vector and a Killing horizon.}

To find the temperature $T_H$ of the Hawking radiation produced because of the presence of the
horizon, we find the surface gravity $k$ from Eq.~(\ref{2}) from the standpoint of an observer with the coordinate
$\rho=\rho_0$. For this, we rewrite this equation in terms of the vector of the ambient space $\xi^a$ given by (\ref{10}).
The covariant derivative of the vector in the embedding theory formalism can be written in the form (this
formalism was presented in \cite{statja18})
\disn{13}{
\na_\al\xi^\m=e^\m_a\dd_\al\ls\xi^\n e_\n^a\rs,
\nom}
where
\disn{14}{
e_\n^a=\dd_\n y^a,\qquad
e^\m_a=g^{\m\n}e_\n^b \eta_{ab}.
\nom}
Multiplying Eq.~(\ref{2}) by $e_\m^b$ and using formula (\ref{13}), we obtain
\disn{15}{
\Pi^b_a\xi^\al\dd_\al\xi^a=k\xi^b,
\nom}
where
\disn{16}{
\Pi^b_a=e^b_\n e^\n_a
\nom}
is the projector onto the space tangent to the surface $\cal M$ at a given point.

For convenience in analyzing Eq.~(\ref{15}), we introduce lightlike coordinates $y^\pm=y^0\pm y^1$ in the ambient
space (analogously to what was done in Sec.~\ref{triv} when discussing the trivial Hawking into Unruh mapping; formulas (\ref{3.4}) and (\ref{3.5})). In these coordinates, formula (\ref{8}) becomes
\disn{17}{
y^+=\rho e^{\al t},\qquad
y^-=-\rho e^{-\al t},\qquad
y^I=y^I(\rho,x^2,x^3),
\nom}
If we use this expression, then we can write formula (\ref{10}) as
\disn{18}{
\xi^+=\frac{\rho}{\rho_0}e^{\al t}=\frac{y^+}{\rho_0},\qquad
\xi^-=\frac{\rho}{\rho_0}e^{-\al t}=-\frac{y^-}{\rho_0}.\qquad
\xi^I=0.
\nom}

It was shown in Sec.~\ref{hyp} that the embedding function for the surface $\cal M$ is smooth
in the coordinates $y^0,y^1,x^2,x^3$ (see~(\ref{11})).
Consequently, it is also smooth in the coordinates $y^+,y^-,x^2,x^3$, in which
it becomes
\disn{19}{
y^+=y^+,\qquad
y^-=y^-,\qquad
y^I=y^I\ls\sqrt{y^+ y^-},x^2,x^3\rs.
\nom}
We write Eq.~(\ref{15}) in these coordinates at the horizon point for which $y^-=0,y^+>0$ (which corresponds
to horizon equation (\ref{11.1})), and this means that $\xi^-=0$ at this point in accordance with (\ref{18}).
For the lefthand
side of Eq.~(\ref{15}), we have
\disn{20}{
\Pi^b_a\xi^\al\dd_\al\xi^a=\Pi^b_a\xi^+\dd_+\xi^a=
\frac{y^+}{\rho_0^2}\Pi^b_a m^a,
\nom}
where $m^a$ is a constant vector with the components $m^+=1;\,m^-=0;\,m^I=0$. It is easy to see
from the form of formula (\ref{19}) that for $y^-=0$, the vector $m^a$ is tangent to $\cal M$
because it coincides with $\dd_+ y^a$. Consequently, $\Pi^b_a m^a=m^b$
in accordance with the property of the projector, and Eq.~(\ref{15}) as a result becomes
\disn{21}{
\frac{y^+}{\rho_0^2}m^b=k\frac{y^+}{\rho_0}m^b.
\nom}
Hence, we can easily find the surface gravity for an observer with the coordinate $\rho=\rho_0$:
\disn{22}{
k=\frac{1}{\rho_0}.
\nom}
This means that the observer in accordance with formula (\ref{n1}) sees Hawking radiation with the temperature
$T_H=1/(2\pi\rho_0)$.

On the other hand, as can be seen from (\ref{8}), this observer in the ambient space moves with the
constant acceleration $w=1/\rho_0$ and consequently sees Unruh radiation with temperature (\ref{2.1}), which turns
out to coincide with $T_H$. The statement is thus proved.

We note that the reasoning used in the proof is in many respects a repetition of that used in Sec.~\ref{triv}
when discussing the trivial Hawking into Unruh mapping. The difference is that
when deriving formula (\ref{22}), we necessarily took the nontriviality of the covariant derivative into account,
and when deriving (\ref{3.7}), the covariant derivative coincided with the ordinary derivative. But the obtained
results were the same in the two cases, i.~e., the Hawking into Unruh mapping
occurs. This can be easily explained as follows.

We note that the difference between general hyperbolic embedding form (\ref{8}) and the expression for
the Minkowski space in terms of Rindler coordinates (\ref{3}) is that the "transverse" components $x^{2,3}$ are
independent of $\rho$ in the Minkowski space and the analogous "transverse" components $y^I$ depend on $\rho$ in
the hyperbolic embedding. But this dependence has no term that is linear in $\rho$ for a smooth embedding
in the case of small $\rho$ (see expansion (\ref{12})). Therefore, in a small neighborhood of the horizon, the smooth
surface $\cal M$ coincides with the Minkowski space for fixed $x^{2,3}$ in the first approximation. For an observer
whose coordinate $\rho$ is small, the Hawking and Unruh temperatures consequently coincide trivially. But for
an observer with an arbitrary value of the coordinate $\rho$, this coincidence is automatically preserved because
of the relation
\disn{23}{
T_U\sqrt{g_{00}}=\frac{1}{2\pi\rho}\sqrt{\al^2\rho^2}=const,
\nom}
which follows from (\ref{2.1}) and (\ref{8}) and is similar in form to the well-known Tolman law
\disn{24}{
T_H\sqrt{g_{00}}=const.
\nom}
The local similarity between such embeddings and the expression for the Minkowski space in terms of the
Rindler coordinates thus underlies the Hawking into Unruh mapping for arbitrary
hyperbolic embeddings smoothly covering the horizon.

We note that we managed to prove the existence of the Hawking into Unruh mapping under rather general assumptions. Concerning the space-time metric, we only assumed the
existence of a Killing timelike vector and a Killing horizon (it is impossible to discuss Hawking radiation
without this) and the existence of a hyperbolic embedding. The embedding function was taken in the most
general form corresponding to the hyperbolic type of realizing the translation invariance under time shifts.
We did not assume the presence of another symmetry, for example, spherical.

\section{Surfaces on which two-point functions\\ are defined by the ambient space}
Quantum effects in a Riemannian space, including the Hawking effect, are defined by the properties of
quantum fields specified in the Riemannian space, for example, by the scalar field satisfying the invariant
equation
\disn{n8.2}{
\ls\na_\m \na^\m+m^2\rs \ff(x)=0.
\nom}
Quantum effects on the embedded surface, in particular, the Unruh effect, are defined by the properties of
quantum fields given in a flat ambient space and satisfying equations of the form
\disn{n8.3}{
\ls\dd_a \dd^a+M^2\rs \ff(y)=0.
\nom}
Equations (\ref{n8.2}) and (\ref{n8.3}) have enough in common that under some additional conditions on the shape of
the surface, a relation between quantum effects in the Riemannian space and on the surface embedded in
the flat ambient space arises.

In addition to the Hawking into Unruh mapping, which was described in the
preceding sections and is such a relation, there is also another analogous relation. It exists for embeddings
for which the ambient space metric that is flat in the Lorentz coordinates $y^a$ becomes the so-called warped
geometry
\disn{n8.1}{
ds^2=\om(z)\bar g_{\m\n}(x)dx^\m dx^\n+g^\p_{AB}(z)dz^A dz^B,
\nom}
in a certain curvilinear system of coordinates $\tilde y^a\equiv(x^\m,z^A)$. Here and hereafter, the superscripts $A,B,\ldots$
take $N-4$ values. It is assumed that the four-dimensional surface corresponding to $z^A=const$ corresponds
to the embedded Riemannian space, and $x^\m$ play the role of coordinates on the surface in terms of which
the metric has the form $g_{\m\n}=\om(z)\bar g_{\m\n}(x)$.
We note that precisely such a method for describing the system
of four-dimensional surfaces -- introducing the $(N-4)$ component field $z^A(y^a)$ -- was used to describe gravity
in the form of splitting theory \cite{statja25} mentioned in Introduction, but the possibility of writing the metric in form (\ref{n8.1})
was not suggested there. On the other hand, we note that when the warped geometry is used in the general
case (for example, in the brane theory framework mentioned in Introduction), it is assumed that an arbitrary
metric rather than the flat metric $\eta_{ab}$ is written in factored form (\ref{n8.1}). This corresponds to a much more
general case than the case considered here where (\ref{n8.1}) is written in the curvilinear coordinates of precisely
a flat metric.

For surfaces corresponding to a factorization of the flat metric in form (\ref{n8.1}), a relation between two-point
Wightman functions on the surface and in the ambient space was established in \cite{hep-th0003098}. This can be
done by separation of variables, as a result of which the quantum field in the ambient space (satisfying
an equation of form (\ref{n8.3})) is expressed in terms of the quantum field satisfying an equation of form (\ref{n8.2})
in the Riemannian space and in terms of a classical variable dependent on the transverse coordinates $z^A$.
Obviously, the possibility of writing the metric in form (\ref{n8.1}) plays a decisive role when obtaining this result,
without which it would be unclear how to separate the variables. We note that dS and AdS spaces were
considered as the ambient spaces in addition to the flat space in \cite{hep-th0003098}.

In \cite{1305.6933,1310.2939} a similar result for the relation between two-point Green's functions was obtained, and
a relation between the causal structure of the ambient space and that of the embedded Riemannian space
was analyzed. The result was also obtained under the assumption that the embedding corresponds to a
factorization of the flat metric of the ambient space in form (\ref{n8.1}).

It is interesting to consider how strong are the restrictions on the structure of the embedded surface
imposed by the condition for metric factorization (\ref{n8.1}) together with the assumption that this metric is flat
(the metric corresponds to the choice of certain curvilinear Coordinates in the flat space). From the standpoint
of the intrinsic geometry, the answer to this question can be obtained as a corollary of Theorem~3.1
in \cite{bertola1}, whence it follows that the embedded surface must be a constant-curvature space. We obtain this
result using the embedding theory formalism \cite{statja18}. In this case, we also managed to simultaneously
find an additional condition constraining the extrinsic geometry.

The metric of the ambient space in curvilinear coordinates $\tilde y^a\equiv(x^\m,z^A)$ is related to the flat metric
$\eta_{ab}$ in the Lorentz coordinates $y^a$ by the standard formula
\disn{n8.4}{
\tilde g_{ab}(\tilde y)=\frac{\dd y^c}{\dd\tilde y^a}\frac{\dd y^d}{\dd\tilde y^b}\eta_{cd},
\nom}
which if (\ref{n8.1}) is taken into account reduces to the equations
\disn{n8.5}{
\ls\dd_\m y^c\rs\ls\dd_\n y_c\rs=\om(z)\bar g_{\m\n}(x),
\nom}\vskip -2em
\disn{n8.6}{
\ls\dd_A y^c\rs\ls\dd_\n y_c\rs=0,\qquad
\ls\dd_A y^c\rs\ls\dd_B y_c\rs=g^\p_{AB}(z),
\nom}
where $\dd_\m\equiv \dd/\dd x^\m$ and $\dd_A\equiv \dd/\dd z^A$. Differentiating equality (\ref{n8.5})
with respect to $z^A$, we obtain
\disn{n8.7}{
\ls\dd_\m\dd_A y^c\rs\ls\dd_\n y_c\rs+
\ls\dd_\n\dd_A y^c\rs\ls\dd_\m y_c\rs=\ls\dd_A \om(z)\rs\bar g_{\m\n}(x).
\nom}
Because the quantity $\dd_A y^c$  with respect to the superscript $c$ is orthogonal to all vectors that are tangent
to the surface $z^A=const$ at a given point in accordance with (\ref{n8.6}), we can move the derivatives in the
left-hand side of (\ref{n8.7}), as a result of which we obtain the equation
\disn{n8.8}{
-2b^c_{\m\n}\dd_A y_c=\ls\dd_A \om(z)\rs\bar g_{\m\n}(x),
\nom}
where
\disn{n8.9}{
b^c_{\m\n}=\na_\m\na_\n y^a=\po^c_a\dd_\m\dd_\n y^a
\nom}
is the second fundamental form of the surface $z^A=const$ (here $\po^c_a=\de^c_a-\Pi^c_a$ is the transverse projector).

Using the inverse metric $g^{\p AB}(z)$ and noting that $(\dd_A y_c) g^{\p AB}(z) (\dd_B y^a)=\po_c^a$,
we can easily rewrite
Eq.~(\ref{n8.8}) in the form
\disn{n8.10}{
b^a_{\m\n}=\xi^a g_{\m\n},\qquad
\xi^a=-\frac{1}{2}(\dd_B y^a)g^{\p AB}(z)\frac{\dd_A \om(z)}{\om(z)},
\nom}
where we take into account that the surface metric is $g_{\m\n}=\om(z)\bar g_{\m\n}(x)$.
Using the expression for the
curvature tensor in the embedding formalism (which is a particular case of the Gauss equation;
see \cite{statja18} for the details), we now obtain
\pagebreak
\disn{n8.11}{
R_{\m\n\al\be}=b^a_{\m\al}\eta_{ab}b^b_{\n\be}-b^a_{\m\be}\eta_{ab}b^b_{\n\al}=
\xi^a\xi_a\ls g_{\m\al}g_{\n\be}-g_{\m\be}g_{\n\al}\rs=\ns=
\frac{\dd_A \om(z)g^{\p AB}(z)\dd_B \om(z)}{4\om(z)^2}
\ls g_{\m\al}g_{\n\be}-g_{\m\be}g_{\n\al}\rs.
\nom}
It hence follows that the surface $z=const$ is necessarily a constant-curvature space, which was to be shown.
But in our argument, we also obtained a stronger condition on the surface that also involves the extrinsic
geometry, i.~e., condition (\ref{n8.10}) on the second fundamental form, in accordance with which it must necessarily
be proportional to the metric.

The simplest variants of surfaces for which this condition is satisfied are the plane, the sphere, and the
pseudosphere (hyperboloid) in an arbitrary number of dimensions. Precisely they are used as embeddings
in a flat ambient space, both in \cite{hep-th0003098} and in \cite{1305.6933,1310.2939}.
But it is not excluded that they are not the only possible
variants. If we use the method proposed in \cite{statja27} for constructing surfaces with given geometries, then we
can try to construct other surfaces whose second fundamental form satisfies condition (\ref{n8.10}). For example,
the nontrivial embedding of the three-dimensional plane has such a property; it was used to construct the
embedding of the metric of the spatially flat Friedmann model in \cite{statja29}. The problem of the properties of
quantum effects for such embeddings requires additional study.

{\bf Acknowledgements.}
The author is grateful to the organizers of the conference "In Search of Fundamental
Symmetries" dedicated to the 90th birthday of Yu. V. Novozhilov.
The work was supported by Saint Petersburg State University grant N~11.38.660.2013.


\end{document}